\documentclass[twoside]{article}
\usepackage{amsmath,amsfonts,amssymb,mathrsfs}
\usepackage{qic,epsfig}
\usepackage{cite}
\usepackage{color}
\usepackage{booktabs}

\newtheorem{thm}{Theorem}
\newtheorem{cor}{Corollary}
\newtheorem{lem}{Lemma}

\renewenvironment{proof}{\par\noindent{\bf Proof.}}{$\quad\Box$\par}
\newcommand{\ket}[1]{| #1 \rangle}
\newcommand{\bra}[1]{\langle #1 |}
\newcommand{\braket}[2]{\langle #1 | #2 \rangle}

\begin{document}
 \date{}

\setlength{\textheight}{8.0truein}    

\runninghead{Algebraic Quantum Error-Correction Codes}
            {}

\normalsize\textlineskip \thispagestyle{empty}
\setcounter{page}{1}

\vspace*{0.88truein}

\alphfootnote

\fpage{1}

 \centerline{\bf Algebraic Quantum Error-Correction Codes}
 \vspace*{0.035truein}
 \centerline{\footnotesize
 Ming-Chung Tsai,
 Po-Chung Chen, Kuan-Peng Chen and Zheng-Yao Su\footnote{Email: zsu@nchc.narl.org.tw; zsu@phys.nthu.edu.tw}\hspace{.15cm}}
\centerline{\footnotesize\it Department of Physics, National Tsing
Hua University, Hsinchu, Taiwan, R.O.C.}
\centerline{\footnotesize\it National Center for High-Performance
Computing, Hsinchu, Taiwan, R.O.C.} \centerline{\footnotesize\it
National Center for Theoretical Sciences, Hsinchu and Tainan,
Taiwan, R.O.C.}

\vspace*{0.21truein}

\abstracts{Based on the {\em group structure} of a unitary Lie
 algebra,
 a scheme is provided to systematically and exhaustively generate quantum error correction codes,
 including the {\em additive} and {\em nonadditive} codes.
 The syndromes in the process of error-correction distinguished by different orthogonal vector subspaces,
 the {\em coset subspaces.}
 Moreover, the generated codes can be classified into four types with respect to the spinors in the unitary Lie algebra
 and a chosen initial quantum state.}{}{}

\vspace*{10pt}  \vspace*{3pt} \vspace*{1pt}\textlineskip

 \section{Introduction}\label{secintro}
 When quantum information is transmitted or manipulated in noisy environments,
 the information gets lost gradually due to the baneful interaction with the environment~\cite{}.
 To protect the fragile quantum states,
 error-correction codes are essential to safeguard the quantum data
 during the processes of quantum computation and communication.
 In this article, a systematic method based on the
 {\em group structure}
 of a unitary Lie algebra $su(2^p)$ as described in~\cite{Su} is introduced to exhaustively generate
 quantum codes, $p\in\mathbb{N}$.
 According to the linking of the group structure in $su(2^p)$
 and admissible quantum codes,
 we are able to construct {\em additive} (stabilizer) quantum error correction codes as well as {\em non-additive} ones.
 Furthermore,
 the generated quantum codes can be classified into four
 types by relating some {\em initial} quantum states and {\em codeword operators}.
 The scheme introduced in this article helps the discovery of new
 types of quantum codes that may have higher efficiency or
 capability to error correction.

 A single qubit state may suffer three types of errors respectively
 represented by the Pauli matrices:
 the {\em bit-flip} error
 $\sigma_1$,
 the {\em phase-flip} error
 $\sigma_3$
 and {\em bit-phase-flip} error
 $\sigma_2$;
 here $\sigma_i$ is defined
 in~\cite{SuTele,Su,SuTsai1,SuTsai2,SuTsai3,SuTsai4}.
 For the $p$-qubit instance, $p\geq 1$, we consider a set of $N$ encountered errors
 ${\cal E}=\{E_0,E_1,\cdots,E_{N-1}\}$
 chosen from the set $G$
 comprising all tensor products of $p$ Pauli matrices
 $E_{\hspace{.5pt}0\leq r<N}=\sigma_{i_1,i_2,\cdots,i_p}\in{G}$.
 A quantum code, denoted as $[[p,K]]$, is a subspace with the code length $p$ and dimension $K$
 of the Hilbert space ${\cal H}_{2^p}$ to protect a piece of information against the above errors.

 \section{Bi-Subalgebra Partition in a Lie Algebra}\label{secSreandBAP}
 To understand the algorithm of constructing quanutm codes,
 the properties of the group structure of the Lie algebra $su(2^p)$
 will be introduced in this section.
 The detailed derivation of these properties of $su(2^p)$ is demonstrated in~\cite{Su,SuTsai1,SuTsai2,SuTsai3,SuTsai4}.
 By writing all the $2^{2p}$ generators (including the identity) of the Lie algebra
 $su(2^p)$ in terms of the spinors in $G$,
 the algebra $su(2^p)$ forms a {\em group under the
 multiplication}~\cite{SuTele,Su,SuTsai1,SuTsai2,SuTsai3,SuTsai4}.
\vspace{6pt}
 \begin{lem}\label{lemsugroup}
 The set of spinor generators of the Lie algebra $su(2^p)$, $p\in\mathbb{N}$,
 forms a group under the multiplication.
 \end{lem}
 \vspace{6pt}
 The most important subgroup of the Lie algebra $su(2^p)$
 is its {\em Cartan subalgebra}.
 A Cartan subalgebra, or called a {\em Cartanion}~\cite{Su},
 $\mathfrak{C}$
 is the {\em maximal abelian subalgebra} of $su(2^p)$.
 It is easy to check that the subalgebra $\mathfrak{C}$
 containing in total $2^p$ generators is a subgroup of $su(2^p)$ under the same group operation~\cite{SuTele,Su,SuTsai1,SuTsai2,SuTsai3,SuTsai4}.
\vspace{6pt}
 \begin{lem}\label{lemmisoC}
 The set of spinor generators of a Cartanion $\mathfrak{C}\subset su(2^p)$ forms an
 abelian group isomorphic to $Z^p_2$ under the mulitplication.
 \end{lem}
\vspace{6pt}

 Being a subgroup of the group $su(2^p)$,
 the Cartanion $\mathfrak{C}$ can generate a {\em partition} in $su(2^p)$,
 which is denoted as $\{\mathcal{P}_{\mathcal{B}}(\mathfrak{C})\}$
 and called the {\em bi-subalgebra
 partition}~\cite{SuTele,Su,SuTsai1,SuTsai2,SuTsai3,SuTsai4}.
 This partition
 $\{\mathcal{P}_{\mathcal{B}}(\mathfrak{C})\}$
 consists of a number $2^p$ of {\em coset subspaces} ${\cal W}_i$
 and satisfies the {\em coset
 rule}~\cite{SuTele,Su,SuTsai1,SuTsai2,SuTsai3,SuTsai4}.
 Note that the Cartanion $\mathfrak{C}$ of course is an element of
 this partition.
 Under this notation,
 the bi-subalgebra partition
 $\{\mathcal{P}_{\mathcal{B}}(\mathfrak{C})\}$
 can form an {\em abelian group} under an appropriate group operation.
 \vspace{6pt}
 \begin{thm}\label{thmiso}
 The bi-subalgebra partition $\{\mathcal{P}_{\mathcal{B}}(\mathfrak{C})\}$
 generated by a Cartanion $\mathfrak{C}$ of the Lie algebra $su(2^p)$
 is isomorphic to the additive group $Z^p_2$.
 \end{thm}
 \vspace{6pt}
 The detailed proof is made constructively in~\cite{Su,SuTsai1}.
 This treatment of partitioning the Lie algebra $su(2^p)$
 into an abelian group structure is essential to the construction
 of error-correction quantum codes in this article.

 \section{Algorithm of Constructing Quantum Codes\label{}}
 A quantum code $[[ p, K]]$ is a subspace of the Hilbert
 space $H_{2^{p}}$ with the code length $p$ and code-subspace dimension $K$, $0< K\leq 2^{p}$.
 The following scheme is given to generate such $[[p,K]]$ codes under the structure of the bi-subalgebra partition
 in $su(2^p)$.
 Firstly, the bi-subalgebra partition
 $\{\mathcal{P}_{\mathcal{B}}(\mathfrak{C})\}=\{{\cal W}_i:i=0,1,\cdots,2^p-1\}$
 generated by a Cartanion $\mathfrak{C}$ in $su(2^p)$
 is prepared.
 Then, one produces the {\em stabilizer} state
 \begin{align}\label{eqinistate}
 \ket{\psi_0}=\sum_{S\in{\mathfrak{C}}}S\ket{\mathbf{0}}
 =\sum^{2^k}_{r=1}(-1)^{\epsilon_r}\ket{\alpha_r}
 \end{align}
 by applying all the spinors of the Cartanion $\mathfrak{C}$
 to the null state $\ket{\mathbf{0}}=\ket{00\cdots 0}$, here $\epsilon_r\in{Z_2}$
 and $\{\alpha_r\}$ being a subgroup of $Z^p_2$.
 Taking the stabilizer state $\ket{\psi_0}$
 as a seed, one can obtain a set of $K$ {\em basis codewords}
 \begin{align}\label{eqbswords}
 BS=\{\ket{\psi_r}:0\leq r<K\}
 =\{\ket{\psi_r}=\hat{S}_r\ket{\psi_0}:0\leq r<K\},
 \end{align}
 each of which is the application of a spinor
 $\hat{S}_r$ in the coset subspace ${\cal W}_r\in\{\mathcal{P}_{\mathcal{B}}(\mathfrak{C})\}$
 to the stabilizer state $\ket{\psi_0}$.
 Here, the set of the $K$ spinors
 \begin{align}\label{eqcodespinor}
 {\cal B}=\{\hat{S}_0,\hat{S}_1,\cdots,\hat{S}_{K-1}\},
 \end{align}
 are respectively chosen from the $K$ different coset subspaces
 and the 1st spinor $\hat{S}_0$ is the identity chosen from
 $\mathfrak{C}$.
 The set $BS$ comprising the $K$ codewords forms a generating set of a code subspace
 $[[p,K]]$ with the length $p$ and dimension $K$.
 Thus as long as a Cartan subalgebra $\mathfrak{C}$ is given,
 there determines the unique partition $\{\mathcal{P}_{\mathcal{B}}(\mathfrak{C})\}$
 such that an enormous number of quantum codes are produced.
 \vspace{6pt}
 \begin{thm}\label{thmCartanQECC}
  Every Cartan subalgebra of the Lie algebra $su(2^p)$ can decide quantum codes
  $[[p,K]]$ with the code length $p$ and dimension $0<K\leq 2^p$.
 \end{thm}
 \vspace{6pt}
 An implication of this theorem is that,
 for a given error set, one can always construct the error-correction codes for this error set by choosing appropriate Cartan
 subalgebra.

 For a given error set
 ${\cal E}=\{E_0,E_1,\cdots,E_{N-1}\}\subset{G}$,
 there exists a
 partition $\{\mathcal{P}_{\mathcal{B}}(\mathfrak{C})\}=\{{\cal W}_{\lambda}:\forall\hspace{2pt}\lambda\in{Z^p_2}\}$
 generated by a chosen Cartan subalgebra $\mathfrak{C}\subset{su(2^p)}$, such that
 the $N$ errors are distributed to $N$ different subspaces,
 namely $E_i\in{\cal W}_{\lambda_i}$ and $\lambda_i\neq\lambda_j$
 if $E_i\neq E_j$,
 here $0\leq i<N$ and $E_0=I^{\otimes p}$.
 The {\em corrupted state}
 \begin{align}\label{eqcorrupt}
 \ket{\psi_{ij}}=E_i\ket{\psi_j}=E_i\cdot S_j\ket{\psi_0}
 \end{align}
 is created by applying the error operator $E_i$ to a basis codeword
 $\ket{\psi_j}=S_j\ket{\psi_0}$, $0\leq i<N$ and $0\leq j<K$.
 We say that the code $[[p,K]]$ has the capability to correct the
 error set ${\cal E}$ if
 \begin{align}\label{eqcondcorrect}
  {\cal W}_{\tau_1}\neq{\cal W}_{\tau_2}\text{ for any }
  E_{i_1}\cdot S_{j_1}\in{\cal W}_{\tau_1}\text{ and }E_{i_2}\cdot S_{j_2}\in{\cal W}_{\tau_2},
 \end{align}
 here $0\leq i_1,i_2<N$, $0\leq j_1,j_2<K$ and
 $\tau_1,\tau_2\in{Z^p_2}$.
 Each corrupted state indicates a {\em syndrome} during the
 process of error-correction and the result of Eq.~\ref{eqcondcorrect}
 means that all the syndromes are distinguishable.
  Thus, the quantum code $[[ p, K]]= span\{ \ket{\varphi_s}\equiv\ket{\varphi_0}, \ket{\varphi_1}, \ket{\varphi_2}, \cdots,
  \ket{\varphi_{K-1}}\}$ and the following lemma is validated.
 \vspace{6pt}
 \begin{lem}\label{maximal}
  Given a stabilizer state $\ket{\varphi_0}$ of a quantum code $[[p, K]]$ decided in the bi-subalgebra partition
  $\{{\cal P}_{\cal B}(\mathfrak{C})\}$ generated by a Cartanion ${\mathfrak{C}}\subset {su(2^p)}$,
  the state $\ket{\tilde{\varphi}_0}=S\ket{\varphi_0}$ acted by a spinor $S\in su(2^p)$ is orthogonal to the stabilizer state
  $\ket{\varphi_0}$, $i.e.,$ $\braket{\tilde{\varphi}_0}{\varphi_0}=0$,
  iff
  $S\notin {\mathfrak{C}}$.
 \end{lem}
 \vspace{3pt}
 \begin{proof}
  From the definition of a stabilizer state $\ket{\varphi_0}$ of
  $[[p, K]]$, the necessity of the condition is obvious by contradiction.
  For the sufficiency,
  considering a maximal bi-subalgebra $\mathfrak{B}$ of $\mathfrak{C}$,
  one has the stabilizer state written as~\cite{Su,SuTsai1,SuTsai2,SuTsai3,SuTsai4}
  $\ket{\varphi_0}= \sum_{S\in\mathfrak{C}} S\ket{\mathbf{0}}
  =\sum_{S_i\in \mathfrak{B}} S_i\ket{\mathbf{0}}
  +\sum_{\bar{S}_i\in \mathfrak{B}^{c}}\bar{S}_i\ket{\mathbf{0}}$.
  The vanishing inner product is obtained
 \begin{align}\label{eqvanishSts}
  \braket{\tilde{\varphi}_0}{\varphi_0}=\bra{\varphi_0}S\ket{\varphi_0}
  =\sum_{S_i\in \mathfrak{B}}
  \bra{\mathbf{0}} S\cdot S_i\ket{\varphi_0}
  -\sum_{\bar{S}_i\in \mathfrak{B}^{c}} \bra{\mathbf{0}}
  S\cdot \bar{S}_i\ket{\varphi_0}
  =0.
  \end{align}
 \end{proof}
 \vspace{6pt}

  The following lemma presents the relationship between the basis codewords generated
  by the stabilizer state and the spinors of the Cartanion.
 \vspace{6pt}
 \begin{lem}\label{codewordeigenvector}
  Given a quantum code $[[ p, K]]= span\{ \ket{\varphi_s}=\ket{\varphi_0}, \ket{\varphi_1}, \ket{\varphi_2}, \cdots, \ket{\varphi_{K-1}}\}$
  decided in the bi-subalgebra partition $\{{\cal P}_{\cal B}(\mathfrak{C})\}$ generated by a Cartanion ${\mathfrak{C}}\subset {su(2^p)}$,
  each basis codeword is an eigenvector of a spinor in $\mathfrak{C}$
  with the eigenvalue $(-1)^\epsilon$ for $\epsilon\in{Z_2}$.
 \end{lem}
 \vspace{3pt}
 \begin{proof}
  A basis codeword $\ket{\varphi_j}$ is an eigenvector of a spinor $S$ in $\mathfrak{C}$
  through the calculation
  $S\ket{\varphi_j}
  =S\cdot S_j\ket{\varphi_0}
  =(-1)^{\epsilon}S_j\cdot S\ket{\varphi_0}
  =(-1)^{\epsilon}\ket{\varphi_j},$
  ${\epsilon}\in {Z}_{2}$, ref.~\cite{SuTele,Su,SuTsai1,SuTsai2,SuTsai3,SuTsai4}.
 \end{proof}
 \vspace{6pt}
  From the quantum code $[[ p, K]]$ and the error spinor set ${\cal E}$,
  the set of {\em syndrome states}
  ${\cal R}_{syn}=\{{\mathbb S}_{ij}\ket{\varphi_0}=\tilde{S}_i\hat{S}_j\ket{\varphi_0}: 0\leq i< N, 0\leq j< K\}$
  is obtained. In other words, each syndrome state
   $\tilde{S}_i\hat{S}_j\ket{\varphi_s}$
  is read as the outcome of the {\em syndrome spinor}
  ${\mathbb S}_{ij}:=\tilde{S}_i\hat{S}_j$, $0\leq i< N$ and $0\leq j < K$,
  applying to $\ket{\varphi_0}$
  and indicates a {\em syndrome} during the process of error-correction.
  Meanwhile, the set of syndrome states generates a {\em syndrome subspace} $H_{syn}$ of $H_{2^{p}}$ with $dim(H_{sym})\leq KN\leq 2^{p}$.

  Nevertheless, to establish a successful quantum error correction code $[[ p, K]]$ to correct an error set $\cal E$,
  it is necessary to distinguish the difference between the syndrome states.
  Lemma~\ref{syndromeeigenvector} expresses
  the relationship between the set of syndrome states and the spinors of the Cartanion extended from
  Lemma~\ref{codewordeigenvector}.
 \vspace{6pt}
 \begin{lem}\label{syndromeeigenvector}
  Given a quantum code $[[ p, K]]= span\{ \ket{\varphi_0}, \ket{\varphi_1}, \ket{\varphi_2}, \cdots, \ket{\varphi_{K-1}}\}$
  decided in the bi-subalgebra partition $\{{\cal P}_{\cal B}(\mathfrak{C})\}$ generated by a Cartanion ${\mathfrak{C}}\subset {su(2^p)}$,
  each syndrome state determined by $[[ p, K]]$ and $\cal E$ is an eigenvector of each spinor of $\mathfrak{C}$.
 \end{lem}
 \vspace{3pt}
 \begin{proof}
  A syndrome state ${\mathbb S}_{ij}\ket{\varphi_0}$
  is an eigenvector of a spinor $S$ in $\mathfrak{C}$
  through the following calculation,
  $S\cdot{\mathbb S}_{ij}\ket{\varphi_0}
  =(-1)^{\epsilon}{\mathbb S}_{ij}\ket{\varphi_0}$
  ${\epsilon}\in {Z}_{2}$,
  where $\pm 1$ is the eigenvalue of $S$ in
  $\mathfrak{C}$ corresponding to ${\mathbb S}_{ij}\ket{\varphi_0}$.
 \end{proof}
 \vspace{6pt}

 \vspace{6pt}
 \begin{thm}
  Given a quantum code $[[ p, K]]= span \{\ket{\varphi_j}: 0\leq j < K \}$
  decided in the bi-subalgebra partition $\{{\cal P}_{\cal B}(\mathfrak{C})\}$
  and an error set ${\cal E}= \{{\tilde{S}}_i: 0\leq i < N \}$,
  two arbitrary syndrome states
  ${\mathbb S}_{ms}\ket{\varphi_0}$
  and ${\mathbb S}_{nt}\ket{\varphi_0}$
  of the syndrome set ${\cal R}_{syn}=\{{\mathbb S}_{ij}\ket{\varphi_0}
  = {\tilde{S}}_i\ket{\varphi_j}, 0\leq i<N, 0\leq j<K \}$
  constructed by $[[ p, K]]$ and $\cal E$ are orthogonal
  $\bra{\varphi_0}{\mathbb S}_{ms}{\mathbb S}_{nt}\ket{\varphi_0}=0$,
  iff
  these code subspaces $\{{\cal W}_{ms}; \forall \ 0\leq m < N, 0\leq s <K, {\cal W}_{00}=\mathfrak{C}\}$
  containing the syndrome spinors respectively are distinguishable,
  $i.e.$, ${\cal W}_{ms}\neq {\cal W}_{nt}$
  for  ${\mathbb S}_{ms}\in{\cal W}_{ms}$
  and $ {\mathbb S}_{nt}\in{\cal W}_{nt}$,
  where $\ket{\varphi_0}$ is a stabilizer state determined by a Cartanion ${\mathfrak{C}}$.
 \end{thm}
 \vspace{3pt}
 \begin{proof}
  The sufficiency of the condition is obvious by Lemma~\ref{maximal}.
  On the other hand, if two arbitrary syndrome states are orthogonal
  $\bra{\varphi_0}{\mathbb S}_{ms}{\mathbb S}_{nt}\ket{\varphi_0}=0$,
  the relations hold,
  ${\cal W}_{ms}\neq {\cal W}_{nt}$
  if ${\mathbb S}_{ms}\neq {\mathbb S}_{nt}$
  by Lemma~\ref{maximal}.
 \end{proof}
 \vspace{6pt}

 \vspace{6pt}
 \begin{cor}\label{nondegenerate}
  Given an error set $\cal E$, there exists a bi-subalgebra partition generated by a Cartanion ${\mathfrak{C}}\subset {su(2^p)}$
  enabling a quantum code $[[ p, K]]$ to correct ${\cal E} \subset {su(2^p)}$.
 \end{cor}
 \vspace{3pt}
 \begin{proof}
  For the set of $N$ errors ${\cal E}=\{\tilde{S}_i:0\leq i<N\}$,
  a such bi-subalgebra partition
  $\{{\cal P}_{\cal B}(\mathfrak{C})\}=\{{\cal W}_m:0\leq m<2^p\}$
  must satisfy the constraint that there exist a number $N$ of
  coset subspaces ${\cal W}_{m_i}$ respectively containing the $N$
  errors $\tilde{S}_i$.
  Together with the $[[ p, K]]=span\{ \ket{\varphi_r}=\hat{S}_r:0\leq r<K\}$,
  the set of syndrome states
  ${\cal R}_{syn}=\{{\mathbb S}_{ir}\ket{\varphi_0}=\tilde{S}_i\cdot \hat{S}_r\ket{\varphi_0}:0\leq i<N,0\leq r<K\}$ is obtained.
  Chose $p$ independent generating spinors as a set
  $\{S_t:1\leq t\leq p\}$ of $\mathfrak{C}$.
  The set syndromes state ${\cal R}_{syn}$ can be ''diagnosed" by these $p$ spinors,
  that is,
  $S_t\cdot{\mathbb S}_{ir}\ket{\varphi_0}=(-1)^{\epsilon_i}{\mathbb S}_{ir}\ket{\varphi_0},$
  ${\epsilon_i}\in {Z}_2$ and $1\leq i\leq p$.
 \end{proof}
 \vspace{6pt}

 \section{Classification of Quantum Codes\label{}}
  Following the steps for constructing quantum codes in the previous section ,
  a basis codeword is created by applying a codeword spinor to a stabilizer state,
  and the corrupted states are consist of the basis codewords and the error set.
  For some special techniques, the four types of quantum codes are constructed as follows.
  Without loss of generality, the classification is generated
  under the quotient-algebra partition determined by a Cartanion.

  Since a Cartanion $\mathfrak{C}$ is a subgroup of $su(2^p)$
  under the multiplication, the set of strings
  $C_{sb}=\{\alpha_m\in{Z^p_2};m=1, 2, \cdots,2^l\}$
  for the stabilizer state
  $\ket{\varphi_s}=\sum_{S_i\in \mathfrak{C}}S_i\ket{00\cdots 0}
  =\sum_{\{\alpha\}\subset Z^{p}_{2}}\ket{\alpha}$
  forms a subgroup of the additive group $Z^p_2$.
  For a fixed stabilizer state $\ket{\varphi_s}$,
  there are two types of quantum codes depending on
  whether the set of codeword spinors ${\cal B}_{cw}$ is a subgroup of $su(2^p)$ under the multiplication.
  To generate the other two types of quantum codes,
  more than one string must be cut off in the set $C_{sb}$ besides the zero string,
  and the stabilizer state $\ket{\varphi_s}$ must not form a subgroup of $Z^p_2$.
  Note that the generated four types of quantum codes by the different options of $\ket{\psi_0}$ and ${\cal B}_{cw}$,
  as shown in Table I.
 \begin{table}[h]
 \begin{center}
 \begin{tabular}{c|cccccc}
  Type && ${\cal B}_{cw}$ && $C_{sb}$& \\
  \hline
  $I$   && g.    && g.    &&  additive\\
  $II$  && n.g.    && g.  &&  nonadditive\\
  $III$ && g.  && n.g.    &&  nonadditive\\
  $IV$  && n.g.  && n.g.  &&  nonadditive\\
 \hline
 \end{tabular}
 \end{center}
  {Table I : The classification of quantum codes generated by the scheme in this paper;
  where g. (n.g.) indicates that ${\cal B}_{cw}$ and $C_{sb}$ are (not) a subgroup of
  Lie algebra $su(2^p)$ and the additive group $Z^p_2$, respectively.} \label{tabclasses}
 \end{table}

  Four types of quantum codes are generated.
  The quantum code of {\em type-I} in Table,
  which is an additive code (stabilizer code),
  corresponds to the choice of $C_{sb}$ and ${\cal B}_{cw}$ being subgroups of $Z^p_2$ and ${\cal P}(\mathfrak{C})$, respectively.
  The remaining codes are nonadditive codes.
  For the code of {\em type-II}, the set ${\cal B}_{cw}$ is not a subgroup, but $C_{sb}$ is.
  The set ${\cal B}_{cw}$ is a subgroup, yet $C_{sb}$ is not for the code of {\em type-III}.
  Neither ${\cal B}_{cw}$ nor $C_{sb}$ are subgroups in the last type of code.

  Both the {\em type-I} and {\em type-II} codes are created by a set
  of codeword spinors ${\cal B}_{cw}$ in $su(2^p)$,
  and a stabilizer state
  $\ket{\psi_0}=\sum_{\{\alpha\}\subset Z^{p}_{2}}\ket{\alpha}$
  whose strings is a subgroup $C_{in}=\{\alpha_m;0\leq m<2^l\}$ in $Z^p_2$.
  Each codeword spinor is included in a subspace
  ${\cal W}_{i}$, $i\in{Z^p_2}$,
  of the partition $\{\mathcal{P}_{\cal B}(\mathfrak{C})\}$ that is generated by a
  Cartanion $\mathfrak{C}\in{su(2^p)}$, {\em cf.} Theorem~\ref{thmiso}.
  Due to the isomorphism of the partition $\{\mathcal{P}_{\cal B}(\mathfrak{C})\}$
  and the additive group $Z^p_2$ by Theorem~\ref{thmiso},
  the behavior of the subspaces ${\cal W}_{i}$ in $\{\mathcal{P}_{\cal B}(\mathfrak{C})\}$
  is equivalent to that of the strings $i$ in $Z^p_2$.
  Since the subspace $\{{\cal W}_{i}\}$ for the quantum code of {\em type-I}
  is a subgroup of $\{\mathcal{P}_{\cal B}(\mathfrak{C})\}$,
  the set of strings $\{i\}$ is also a subgroup of $Z^p_2$.
  Thus, this type of quantum code has a {\em linear} correspondence in the classical regime.
  Since the subspaces $\{{\cal W}_{i}\}$ for the code of {\em type-II} is not a subgroup of $\{\mathcal{P}_{\cal B}(\mathfrak{C})\}$,
  the set of strings $\{i\}$ is not a subgroup of
  $Z^p_2$.
  Thus, this type of quantum code has a {\em nonlinear} correspondence
  in the classical regime.

 \section{Conclusion}
  Based on the group structure of the Lie algebra $su(2^p)$,
  this study presents a scheme to generate an exhaustive set of quantum codes $[[p,K]]$
  with code length $p$ and dimension $1\leq K\leq 2^p$ systematically.
  The study further classifies these generated quantum codes into
  four types according to the relations of codeword spinors and a given stabilizer quantum state.
  New types of quantum codes can be discovered for the purpose of
  searching the quantum codes higher efficiency and capability for error correction.

\nonumsection{References}

\end{document}